\documentstyle[12pt]{article}

\newcommand{\beq}{\begin{equation}}
\newcommand{\eeq}{\end{equation}}
\newcommand{\beqa}{\begin{eqnarray}}
\newcommand{\eeqa}{\end{eqnarray}}
\newcommand{\beqan}{\begin{eqnarray*}}
\newcommand{\eeqan}{\end{eqnarray*}}

\newcommand{\ul}{\underline}
\newcommand{\ol}{\overline}

\newcommand{\ben}{\begin{enumerate}}
\newcommand{\een}{\end{enumerate}}
\newcommand{\bfl}{\begin{flushleft}}
\newcommand{\efl}{\end{flushleft}}
\newcommand{\ba}{\begin{array}}
\newcommand{\ea}{\end{array}}
\newcommand{\btab}{\begin{tabular}}
\newcommand{\etab}{\end{tabular}}
\newcommand{\bit}{\begin{itemize}}
\newcommand{\eit}{\end{itemize}}

\newcommand{\hs}{\hspace}

\def\beq{\begin{equation}}
\def\eeq{\end{equation}}

\newcounter{muni}

\begin{document}
\pagestyle{empty}
\thispagestyle{empty}
%\voffset=.5truein

%\baselineskip 16pt plus 2pt minus 2pt
%\vglue 1.5truein
%\begin{titlepage}
\begin{flushright}
BROWN-HET-1007

BROWN-TA-525

%hep-ph/9511xxx
\end{flushright}
\vspace{5mm}

\begin{center}
%{\Large \bf Remarks on the UA4/2 d$\sigma$/dt \\
%and on the Asymptotic $\sigma_{T}(s)$ Behavior \\}

{\Large \bf Remarks on the UA4/2 d$\sigma$/dt \\
and on the Asymptotic $\sigma_{T}(s)$ Behavior\footnote{Supported in part
by the USDOE Contract DE-FG02-91ER 40688-Task A and presented at the
International Conference (VIth Blois Workshop) on Elastic and Diffractive
Scattering, 20-24 June 1995, Chateau de Blois, France} \\}
\vspace{3mm}
{\bf Kyungsik Kang  \\
Department of Physics, Brown Unversity, Providence, RI 02912, USA} \\
%\end{center}

%\vs{1mm}

%\vskip 1truein

%\vs{2mm}
\centerline{and}
%\vs{2mm}
{\bf Pierre Valin \\
Labo. de Physique Nucleaire, Universite de Montreal, Montreal,
Quebec, Canada}

%\vs{2mm}

%\vs{2mm}

\begin{abstract}
{
We find that the UA4/2 measurements on the real part of the forward
$p \bar p$ scattering amplitude and total cross section are consistent
with each other and also with the standard picture of the Pomeron of
either ${\it ln} ^2 s$ or ${\it ln} s$ type. However the asymptotic
$\sigma _T (s)$ behavior obtained from the {\it ad hoc} parametrization
fit to the $p \bar p$ total cross sections as quoted in the Review of
Particle Properties is consistent with the analysis of the class of
the analytic amplitude models that contain the ${\it ln} ^2 s$ type
Pomeron term.  }
\end{abstract}
\end{center}
%\end{titlepage}
%\vs{2mm}
%\newpage
\noindent
{\bf I. Preamble}
\\
\indent
Let us first recall a few definitions. The $pp$ and $\ol{p}p$
elastic scattering amplitudes can be decomposed into their C-even and
C-odd components by
$ F_{PP} = F_{+} + F_{-} $ and $ F_{\ol{P}P} = F_{+} - F_{-} $
or equivalently,
$ F_{\pm} =  ( F_{PP} \pm F_{\ol{P}P} ) / 2 $.
The  total cross section is given by the optical theorem
$ \sigma_{T}(s) = { (1 / s)} \hs{2mm} Im F(s, t = 0)$
and the forward ratio parameter is defined by
$$ \rho(s) = {Re F(s, t=0) \over Im F(s, t = 0)}        \eqno(1)$$
\noindent
In terms of such amplitudes, the differential cross section reads
$${d\sigma \over dt} = {1 \over 16 \pi s^2 (\hbar c)^2} | F(s, t) |^2
                                                        \eqno(2)$$
\noindent
The differential cross sections for $pp$ and $p\ol{p}$ scattering
can then be separated into Coulomb and nuclear components
$${d\sigma \over dt} ={1 \over 16 \pi s^2 (\hbar c)^2} | F_C + F_N
|^2                                                      \eqno(3)$$
where the Coulomb part is related to the electric charge form factor
by the usual formulae$^{[1]}$
$$ F_C = {8 \pi \alpha (\hbar c)^2 s \over |t|}
G^2(t) \hs{2mm} e^{-i \alpha \phi(t)}                    \eqno(4)$$
with the Coulomb phase approximately given by
$ \phi(t) = ln \left({0.08 / |t|} \right) - 0.577$
and the electric charge form factor parametrized by its dipole form
$ G(t) = \left( 1 + { |t| / 0.71} \right)^{-2} $.
For sufficiently small angles, the hadronic amplitude obeys an
exponential form
$$ F_N(s, t) = F(s, t) = s \hs{2mm} \sigma_{T}(s) \hs{2mm}
( \rho(s) + i ) \hs{2mm} e^{- {1 \over 2} B |t|}          \eqno(5)$$
%These formulae can be used to rewrite
Sometimes the differential cross sections are written
in a simpler form by changing normalizations such that
%$${d\sigma \over dt} = \pi \left| f_C + f_N  \right|^2     \eqno(6)$$
 $ f_C = F_C /(4 \pi \hbar c s)$ etc.
%$  f_C = {2 \alpha (\hbar c)^2 \over |t|}
%G^2(t) e^{-i \alpha \phi(t)} and
%\hs{3mm}, \hs{3mm}
%f_N = {\sigma_{tot} \over 4 \pi \hbar c}
%(\rho + i) e^{- {B \over 2} t} $.
%A certain functional combination of the total cross-section
%and the ratio parameter can be obtained by a luminosity ($L$)
%independent method.
Given that the total cross section can be
separated in elastic and inelastic counts
%\beqa
$\sigma_{T} = {(1 / L)} \hs{2mm} ( R_e + R_{i} ), \hs{5mm}
%& & %\no
{d\sigma / dt} = {(1 / L)} \hs{2mm} {(dR_e / dt)} $, \hs{2mm}
%& & \no \\
%\hs{5mm}
we obtain
$$\left( {d\sigma \over dt} \right)_{t = 0} = \hs{2mm}
{1 \over 16 \pi s^2 (\hbar c )^2 }
\hs{2mm} | F_N |^2_{t =0}
%\no \\& & \hs{40mm}
 \hs{2mm} = \hs{2mm} {1 \over 16 \pi s^2 (\hbar c )^2 } \hs{2mm}
\sigma_{T}^2 ( 1 + \rho^2)                               \eqno(6)$$
%\no
%\eeqa
$$ \sigma_{T}^2(s) = {16 \pi (\hbar c)^2 \over 1 + \rho^2} \hs{2mm}
{1 \over L } \hs{2mm}  {dR_e \over dt} |_{t = 0}         \eqno(7)$$
so that
$$ \sigma_{T}(s) ( 1 + \rho^2) = 16 \pi (\hbar c)^2 \hs{2mm}
{ \left[ {dR_e / dt} \right]_{t=0} \over R_e + R_i } \eqno(8)$$
which is the well-known L-independent expression. Data for $\ol{p}p$
on Eq. (8) comes both from CERN's UA4 Collaboration$^{[2]}$
at $\sqrt{s} = 546$ GeV, $  \sigma_{T}(s)(1 + \rho^2(s)) = 63.3 \pm 1.5
\hs{2mm}mb $, and from Fermilab's CDF Collaboration$^{[3]}$
at $\sqrt{s} = 546$ GeV,  $\sigma_{T}(s)(1 + \rho^2(s)) = 62.64 \pm 0.95
\hs{2mm}mb $ and at $\sqrt{s} = 1800$ GeV,
$\sigma_{T}(s)(1 + \rho^2(s)) = 81.83 \pm 2.29 \hs{2mm}mb $.
A complete set of total cross section and real part data compilation, including
statistical merging of data points at a given energy and also updated with all
existing Tevatron data, can be found elsewhere$^{[4, 5]}$.

There are two recent elastic scattering data from the UA4/2 collaboration:

(1). $\rho =  0.135 \pm 0.015$	and $B =  15.5 \pm 0.2 \hs{2mm}(GeV/c)^{-2}$ at
$\sqrt{s} =  541 \hs{2mm}GeV ^{[1]}$
in the interval $0.875 \cdot 10^{-3} \leq |t| \leq 0.1187 \hs{2mm}(GeV/c)^2$
subject to the UA4  L-independent result
$(1 + \rho^2) \hs{2mm} \sigma_{T}(s) = 63.3 \pm 1.5 \hs{2mm}mb$
at $\sqrt{s} = 546 \hs{2mm}GeV$, and

(2). The L-dependent $\sigma_{T}(s)$ determination$^{[6]}$
$\sigma_{T}(s) = 63.0 \pm 2.1 \hs{2mm} mb$, which can be compared to
$\sigma_{T}(s) = 62.2 \pm 1.5 \hs{2mm}mb$
obtained from item (1). Similarly, from the $\rho$ value of item (1) this total
cross section gives $(1 + \rho^2) \hs{2mm} \sigma_{T}(s) = 64.15 \hs{2mm}mb$.
\\
\noindent
{\bf II. $\rho$, $B$ and $\sigma_{T}$ from UA4/2 Experiment}
\\
\indent
We should note that the UA4/2 $\rho$ is consistent with the standard
picture of the Pomeron dominance (either $ln^2 s $ or $ln s$ type
extrapolations) and thus there is little room for non-standard type of new
physics$^{[4, 5]}$. Nevertheless, the reason behind reexaming $d\sigma /dt$
is to see if any combination of the following inputs to Coulomb fits can
tolerate
or even suggest alternatives to the standard picture :

(1). Dipole vs. other form factors: Felst, BSWW, etc.$^{[7]}$ made very little
changes. \\
(2). Choice of $t$ regions: But one must be careful to include enough of small
$t$
(Coulomb peak) and of large $t$ data (to show the nuclear slope consistent
with previous measurements). We selected two sets:
$t = 0.875 \cdot 10^{-3}$ to $0.395 \cdot 10^{-1} \hs{2mm}(GeV/c)^2$ for
a total of 67 points (medium t range); and
$t = 0.875 \cdot 10^{-3}$ to $0.11875 \hs{2mm}(GeV/c)^2 $ for a grand
total of 99 points (full t range).

We find that these ranges affect somewhat the results,
particularly the size of the parameter errors as explained below.

(3). Sensitivity of assuming that $\sigma_{T}(s)$ is given independently
or only through the combination $(1 + \rho^2) \hs{2mm} \sigma_{T}$:
To study this, we first assume $\sigma_{T}(s) = 63.0 \pm 2.1 mb$
at $\sqrt{s} = 541 \hs{2mm}GeV$ in the UA4/2 experiment
independently of $(1 + \rho^2) \hs{2mm} \sigma_{T}(s)$ and fit the UA4/2
t-distribution for the two t-ranges of item 2.  The results of fits
are as following:
\\
\noindent
{\bf \ul{medium t range} : \hs{50mm} \ul{all t range} : } \\
 \noindent
$\sigma_{T} (mb)$ \hs{2mm} $\chi^2$ \hs{3mm} $\rho(s)$ \hs{20mm} $B
(GeV/c)^{-2}$
\hs{3mm}   $ \sigma_{T} (mb)$ \hs{2mm} $\chi^2$ \hs{3mm} $\rho(s)$ \hs{20mm}
$B (GeV/c)^{-2}$  \\
60.9  \hs{2mm} 77.75 \hs{2mm} 0.129 $\pm$ 0.013 \hs{2mm} 15.326 $\pm$
0.205 \hs{3mm}60.9  \hs{2mm} 110.86 \hs{2mm} 0.118 $\pm$ 0.008 \hs{2mm} 15.546
$\pm$ 0.061 \\
63.0  \hs{2mm} 69.95 \hs{2mm} 0.175 $\pm$ 0.014 \hs{2mm} 15.084 $\pm$
0.207 \hs{3mm} 63.0  \hs{2mm} 106.84 \hs{2mm} 0.153 $\pm$ 0.009 \hs{2mm} 15.484
$\pm$ 0.061  \\
65.1  \hs{2mm} 70.28 \hs{2mm} 0.222 $\pm$ 0.016 \hs{2mm} 14.870 $\pm$
0.208 \hs{3mm} 65.1  \hs{2mm} 112.09 \hs{2mm} 0.188 $\pm$ 0.010 \hs{2mm} 15.428
$\pm$ 0.061  \\
\indent
A few comments are in order about these fits: \\
\noindent
$-$ $\sigma_{T}$ increases as $B$ decreases and as $\rho$ increases indicating
a negative $\rho - B$ correlation (as reported by UA4/2) !
The strengths of the correlation are, however, dependent of the t-range. \\
\noindent
$-$ For medium t, $\chi^2$ rises much faster at the lower end than at the
higher
end of $\sigma_{T}(s)$, while for all t, $\chi^2$ changes slowly and
symmetrically !
As expected, the smaller t-range correlates with larger parameter errors.
Also for medium t, $B$ tends to be too far from data set for higher total
cross-sections (i.e., recall that $B = 15.5 \pm 0.2 \hs{2mm}(GeV/c)^{-2}$ for
UA4/2 and $B = 15.28 \pm 0.58 \hs{2mm}(GeV/c)^{-2}$ for CDF). Any variation of
$F_N (s, t)$ from Eq.(5) with more complicated t-dependence$^{[7, 8]}$ must be
tested with all $t- \rho $ range data to insure $\chi^2$ stability. \\
\noindent
$-$ For all t, $\sigma_{T} = 63.0 \hs{2mm}mb$ and $\rho = 0.153$,
which corresponds to $(1 + \rho^2) \hs{2mm} \sigma_{T} = 64.475 \hs{2mm}mb$,
i.e., within 1 $\sigma$ of $63.3 \pm 1.5 \hs{2mm}mb$. But $\rho = 0.153 \pm
0.009$
is at 2 $\sigma$ of $\rho = 0.135$ or equivalently the experimental value
$\rho = 0.135 \pm 0.015$ is at 1.2 $\sigma$ of $\rho = 0.153$. Two fits of the
t-distribution are shown for the two different t-ranges in Fig. (1)
for comparison. \\
\noindent
$-$ Finally, the case in which $\rho \simeq 0.135$ and
$(1 + \rho^2) \hs{2mm} \sigma_{T} = 63.3 \hs{2mm}mb$ ( so that
$\sigma_{T}(s) = 62.2 \hs{2mm} mb$) is perfectly consistent with our
interpolation of $\rho$ from the above.
In fact, for all t, we get for $\sigma_{T} = 62.0 \hs{2mm}mb$,
$\rho = 0.136 \pm 0.009$, $B = 15.512 \pm 0.061 \hs{2mm}(GeV/c)^{-2}$ and
$\chi^2 = 107.45$.

(4). Sensitivity of fast slope changes:
The allowed slopes from the medium and full t-range vary little,
leaving little room for such variations in model fits$^{[8]}$. This is clear
also if one assumes $(1 + \rho^2) \hs{2mm} \sigma_{T}(s) = 63.3 \pm 1.5
\hs{2mm}mb$ at $\sqrt{s} = 546 \hs{2mm} GeV$, for then one obtains for all
t-range the following fit: \\
\noindent
$(1 + \rho^2)\sigma_T (mb)$ \hs{5mm} $\chi^2$ \hs{28mm} $\rho(s)$
\hs{15mm} $B (GeV/c)^{-2}$ \hs{15mm} $\sigma_T (mb)$ \hfill\break
\hs{5mm} 63.3  \hs{20mm} 107.25 \hs{15mm} 0.137 $\pm$ 0.007 \hs{5mm}
15.512$\pm$ 0.058 \hs{15mm} 62.13 \\
\hs{5mm} 64.8  \hs{20mm} 107.01 \hs{15mm} 0.157 $\pm$ 0.007 \hs{5mm}
15.475$\pm$ 0.058 \hs{15mm} 63.23 \\
\noindent
{\bf III. Asymptotic $\sigma_T(s)$}
\\
\indent
 The 1994 Review of Particle Properties$^{[9]}$ quotes the fits
for a number of hadronic total cross sections to the {\it ad hoc}
parametrization
$$\sigma_T = A + B \hs{2mm} p^n + C \hs{2mm} ln^2 p + D \hs{2mm} ln p
                                                          \eqno(9)$$
where $p$ = the beam momentum in GeV/c. However such a parametrization has no
theoretical basis and furthemore it is difficult to give physical
interpretations
to the parameters, nor to provide any correlation from
reaction to reaction !

On the other hand, the analytic amplitude models$^{[4, 5]}$ based on the
general principles can give natural physical interpretations to the parameters.
We may then regard the
analytic amplitude models as representing an "average" of the
pure empirical parametrization, Eq. (9),
at high energies. In particular we may extract the asymptotic
$\sigma_T(s)$ behavior of Eq. (9), i.e.,
$\sigma_T \propto C \hs{3mm} ln^2 s + d \hs{3mm} ln s + .....$
from their result and compare it with the results of the analytic amplitude
model fits$^{[5]}$. This can be done most appropriately for the $p\bar p$
reaction which has the most data at high energies.
For $\ol{p}p$, the revised CERN-HERA and COMPAS fits$^{[9]}$ give
$C = 0.26 \pm 0.05$ and $D = - 1.2 \pm 0.9$.

The class of models that can be compared to their fits at high energies are
either $P_2 + O + \Sigma Regges$ or $P_2 + \Sigma Regges$ types where
$C = B_{+}$  unambiguously and $D = 2B_{+} (ln (2m) - ln s_{+}) + \pi B_{-}$
in the notations of Ref.[5]. The models with the $ln s$-type Pomeron term give
$C = 0$ even though the $P_1 + R_D + R_{ND}$ model has the most preferred
$\chi^2/d.o.f$ value, 1.30 for $\sqrt {s} \geq 9.7$ GeV. The results are:

\noindent
$C = B_{+} = 0.2425, \hs{5mm} D = -0.1155$ \hs{10mm} for the $P_2 + R_D +
R_{ND}$
model ($\chi^2/d.o.f = 1.32 $), \\
\noindent
$C = B_{+} = 0.2328, \hs{5mm} D = -0.3273$ \hs{2mm} for the $P_2 + O + R_D +
R_{ND}$
model ($\chi^2/d.o.f = 1.35$), \\
\noindent
$C = B_{+} = 0.2301, \hs{5mm} D = -0.0961$  \hs{7mm}for the $P_2 + O + R_{ND}$
model
($\chi^2/d.o.f = 1.33$, and \\
\noindent
$C = B_{+} = 0.2279, \hs{5mm} D = 0.1057$  \hs{25mm} for the $P_2 + R_{ND}$
model
($\chi^2/d.o.f = 1.37$).
\\
\noindent
{\bf IV. Conclusions}

Concerning the UA4/2 $\rho$ and $\sigma_T$, we find that:
\\
\noindent
$-$ the measurements are consistent with the standard picture of the
Pomeron (either $ln^2 s $ or $ln s$) and little room for the non-standard new
physics, and
\\
$-$ independently of assuming $\sigma_T$ through $(1 + \rho^2) \sigma_T =$
fixed,
one can reproduce the UA4/2 $\rho$, $B$ and $\sigma_{T}$ from their
dN/dt data, i.e., ($\rho = 0.136 \pm 0.009$, $B = 15.512 \pm 0.061 \hs{2mm}
(GeV/c)^{-2}$, $\sigma_T = 62.0 \hs{2mm}mb$) or ($\rho = 0.157 \pm 0.007$,
$B = 15.475 \pm 0.058 \hs{2mm}(GeV/c)^{-2}$, $\sigma_T = 63.23 \hs{2mm}mb$).

Concerning the asymptotic $\sigma_T(s)$ for $p\bar p$, we find that:
\\
\noindent
$-$ the analytic amplitude models with $ln^2 s$ type Pomeron term give a
consistent
asymptotic $\sigma_T (s)$ behavior with that of the {\it ad hoc}
parametrization fit by
the CERN-HERA and COMPAS groups, i.e., $C = B_{+} = 0.23 \sim 0.24$ and
$D = -0.33 \sim 0.11$.
\\
%\endpage
\noindent
{\bf References}
%\begin{description}
\\
{\bf 1.} UA4/2 Collab., C. Augier et al., Phys. Lett. {\bf B 316} (1993) 448.\\
{\bf 2.} UA4/2 Collab., M. Bozzo et al., Phys. Lett. {\bf B 147} (1984) 392.\\
{\bf 3.}
CDF Collab., F. Abe et al., Phys. Rev. {\bf D 50} (1994) 5550. \\
{\bf 4.} K. Kang, P. Valin and A. White, in Proc. Vth Blois Workshop (1993);
Nuovo Cimento \\
\indent
{\bf 107A} (1994) 2103.  \\
{\bf 5.} K. Kang and S.K. Kim, in these Proceedings. \\
{\bf 6.} UA4/2 Collab., C. Augier et al., Phys. Lett {\bf B 344} (1995) 451.\\
{\bf 7.}
R. Felst, Desy Report 73/56 (Nov. 1973); F. Borkowski, G. G. Simon, V. H.
Walther and \\
\indent
 R. D. Wendling, Nucl. Phys. {\bf B93} (1975) 461. See also
M. M. Block, in these Proceedings. \\
{\bf 8.} O. V. Selyugin, Phys. Lett. {\bf B 333} (1994) 245; in these
Proceedings.
See also V. Kundrat, \\
\indent
 in these Proceedings. \\
{\bf 9.} Particle Data Group, L. Montanet et al., Phys. Rev. {\bf D50} (1994)
1173.
See p. 1335. \\
%\end{description}
\noindent
{\bf  Figure Captions} \\
%\begin{description}
{\bf Fig. 1} {\bf (a)}. $d\sigma /dt$ for medium
 t and {\bf (b)}. for all t  when
$\sigma_T = 63.0 mb$ is assumed in the UA4/2 experiment independently of the
UA4 result on $(1 + \rho ^2) \sigma_T (s)$. \\
%\end{description}
\end{document}